\documentclass[useAMS,usenatbib]{mn2e}
\usepackage{aas_macros}
\usepackage{amsmath}
\usepackage{epsfig}
\usepackage{graphicx}

\def\lesssim{\lower.5ex\hbox{$\; \buildrel < \over \sim \;$}}
\def\gtrsim{\lower.5ex\hbox{$\; \buildrel > \over \sim \;$}}

\def\msolar{M$_\odot$}

\def\jcap{JCAP}

\newcommand{\na}{New Astronomy}

\newcommand{\msun}{\mbox{${\rm M}_{\odot}$}}

\newcommand{\mum}{\mbox{$\mu$m}}
\newcommand{\lya}{Lyman-$\alpha$}
\def\lesssim{\lower.5ex\hbox{$\; \buildrel < \over \sim \;$}}
\def\gtrsim{\lower.5ex\hbox{$\; \buildrel > \over \sim \;$}}
\newcommand{\snpi}{SN$_{\mbox{\scriptsize PI}}$}

\title[Constraining the near-IR background light]{Constraining the near-IR background light from Population-III stars using high redshift gamma-ray sources}

\author[R.~C.~Gilmore]{Rudy C. Gilmore$^{1}$\thanks{E-mail: rgilmore@sissa.it}\\
$^{1}$SISSA, via Bonomea 265, 34136 Trieste, Italy
}

\begin{document}

\date{\today}

\pagerange{\pageref{firstpage}--\pageref{lastpage}} \pubyear{2011}

\maketitle
\label{firstpage}

\begin{abstract}
The {\em Fermi} satellite has detected GeV emission from a number of gamma-ray bursts and active galactic nuclei at high redshift, $z \gtrsim 1.5$.  We examine the constraints that the detections of gamma rays from several of these sources place on the contribution of population-III stars to the extragalactic background light.  Emission from these primordial stars, particularly redshifted \lya~emission, can interact with gamma rays to produce electron-positron pairs and create an optical depth to the propagation of gamma-ray emission, and the detection of emission at $>$10 GeV can therefore constrain the production of this background.  We consider two initial mass functions for the early stars, and use derived SEDs for each to put upper limits on the star-formation rate density of massive early stars from redshifts 6 to 10.  Our limits are complementary to those set on a high near-IR background flux by ground-based TeV-scale observations, and show that current data can limit star-formation in the late stages of reionization to less than 0.5 \msolar~yr$^{-1}$ Mpc$^{-3}$.  Our results also show that the total background flux from population-III stars must be considerably less than that from resolved galaxies at wavelengths below 1.5 \mum. 
\end{abstract}


\section{Introduction}

The reionization of the universe, which took place around a redshift of 10.5 \citep{komatsu11}, is generally believed to be driven primarily by ionizing photons from early `Population III' (pop-III) stars.  As these stars form from primordial unenriched, metal-free hydrogen and helium, they undergo a formation process that is substantially different from that of later population I and II stars.  Simulations of the production of pop-III stars \citep{abel02,bromm02,tan&mckee04,yoshida06,norman08} generally find an initial mass function (IMF) that is heavily biased towards high masses, 10 -- 1000 \msolar.  A considerable fraction of the radiant energy from these stars is released at ionizing wavelengths ($< 912$ \AA),  which allows reionization of the universe to be completed on the timescale required by \lya~forest data \citep{becker01}.  Despite their importance in cosmology and impact on IGM evolution, pop-III stars continue to evade direct detection.  Detecting the redshifted UV emission from pop-III stars is a primary goal of the upcoming {\em James Webb Space Telescope}, though even with the state of the art sensitivity of this instrument detecting individual metal-free stars will be challenging \citep{rydberg11}.  Searching for indirect evidence of these stars and their integrated cosmological impact is therefore the primary way of understanding the properties of the reionization-era universe.

Photon production from the reionization era is encoded in the evolving spectral energy distribution (SED) of the accumulated photon background, which we observe locally as the extragalactic background light (EBL).  Redshifted UV radiation from pop-III stars can be expected to appear as a distinct component of the near-IR portion of the EBL, and the spectral details of this observable light could in principle inform an observer about the redshift of reionization and the nature of the sources responsible \citep{kashlinsky04,cooray04,madau&silk05,fernandez&komatsu06}.  However, observations of the absolute intensity of the EBL in the near-IR are severely hindered by the presence of bright galactic foregrounds, which are produced by diverse sources including stars and the interstellar medium (ISM) of the Milky Way and `zodiacal light' from dust within our solar system \citep{hauser&dwek01}.  

It has been proposed that a high level of IR background could be due to radiation from the first generation of stars. Multiple experiments, most notably the DIRBE experiment on the COBE satellite, have attempted to measure the sky brightness at near-IR wavelengths, and foreground subtraction analyses have been presented by a number of authors \citep{wright&reese00,wright01,gorjian00,cambresy01,levenson07}, with extragalactic components generally exceeding the contribution of resolvable galaxies.  As discussed in \citet{gspd11} (GSPD11), the discrepancies between these various measurements are largely due to the models for the impact of zodiacal light assumed.  The IRTS experiment on the Space Flyer Satellite found an even higher measurement than DIRBE in the near-IR at 1.4 to 4 \mum~\citep{matsumoto05}, and linked this radiation excess to pop-III formation.  

Other findings have cast doubt on this interpretation of the high near-IR background flux arising from reionization-era sources.    TeV-scale observations by ground-based atmospheric Cherenkov telescopes can constrain the EBL at wavelengths from the UV to the far-IR \citep{dwek&krennrich05,aharonian06,mazin&raue07,albert08} through the optical depth introduced by pair-production interactions with background photons to gamma rays from extragalactic sources such as gamma-ray emitting blazars.  Constraints of this type, which assume a limit to the intrinsic hardness of gamma-ray spectra, have generally found agreement with a low value for the EBL in the near-IR.  While these limits can be violated in some circumstances, a EBL flux as high as that measured by the IRTS experiment and the higher DIRBE claims would almost certainly not be possible without the emergence of new physics.  Moreover, \citet{madau&silk05} (hereafter MS05) investigated the theoretical implication of the claim of a large IR background excess, and found that creating a near-IR background excess would be inconsistent with the observed chemical enrichment history of galaxies.

In this work, we revisit the possibility of constraining the contribution of reionization-era stars to the EBL using gamma-ray observations, and we then address the corresponding limits on high-redshift star formation.  Rather than focusing on the relatively low-redshift blazars that have been viewed with ground-based gamma-ray telescopes and that have been the target of aforementioned EBL studies, here we explore the implications of the most distant sources observed by the Large-Area Telescope (LAT) on the {\em Fermi} Satellite at lower GeV energies.  The gamma-ray optical depth for high-redshift blazars and gamma-ray bursts (GRBs) due to the EBL produced by galaxies in a variety of models was the topic of a recent {\em Fermi} collaboration paper \citep{fermiEBL}.   However, none of the EBL models discussed in this paper included explicitly a contribution from the high-redshift UV sources responsible for reionization, like pop-III stars.  Here, we will employ a similar method to \citet{fermiEBL}, by using the highest energy photons seen from several high-redshift sources, together with the spectrum observed at lower energies, to put an upper limit on the background flux, and we treat the flux from pop-III stars as a separate component from the EBL produced by later galaxies.  Our work is also related to the calculation of \citet{raue09} (RKM09), which examined the limits possible on low- and zero-metallicity stars from gamma-ray data and placed limits on reionization-era star formation from redshifts 7 to 15.  This analysis was limited to considering the EBL bounds placed by comparatively low-redshift TeV blazars, rather than the high redshift GeV sources we consider in this work.

The prospects of using high redshift observations with {\em Fermi}, in particular of distant gamma-ray bursts, as probes of light originating in the reionization era was noted in \citet{kashlinsky05b} and \citet{kashlinsky&band07}.  These papers recognized that the gamma-ray optical depth introduced by photons from early stars would create a $(1+z)^{-3}$ dependence in the gamma-ray mean free path, due to the constant comoving number density of the resulting background photons, and that the highest-redshift sources would therefore be the most sensitive to the spectral signature of this radiation.   High redshift gamma-ray spectra, when used as a probe of background photons, can therefore effectively isolate an EBL component originating in the early universe. Though the LAT has been less successful at detecting high-energy GRBs than envisioned by these authors, high-redshift detections of both GRBs and blazars have been made at energies above 10 GeV, and these are sufficient to constrain a large excess at the shorter bands of the near-IR wavelengths.

In the next section, we first review the calculations of the buildup of background photons and the phenomenon of gamma-ray attenuation through electron-positron pair-production, and then describe a model for the SED of the background component of pop-III stars.  In Section \ref{sec:results}, we present our results for the limits that can be obtained from observed gamma-ray sources on the portion of the local EBL that originates from high-redshift, followed by upper bounds on the total rate of pop-III star formation.  Section \ref{sec:disc} presents conclusions and the prospects of strengthening our limits with future high-redshift gamma-ray source detections.  We assume a $\Lambda$CDM cosmology with parameters consistent with a maximum likelihood results of \citet{komatsu11} in this work: $h=0.702$, $\Omega_b = 0.0455$.

\section{Methods}

To examine the effect that the photon populations created during reionization have on gamma-ray propagation, we will develop template SEDs that describe the emissivity of the universe during these early times.  These templates can be freely rescaled to simulate varying star-formation rate densities.  They are described in full detail in \S \ref{sec:spectmod}.

In the following sections, we will treat independently the hypothetical EBL component from reionization-era pop-III stars and the component produced primarily by later population I/II stars.  We will distinguish between these by referring to the first as the reionization-era EBL (`r-EBL') and the second as the post-reionization EBL (`p-EBL').  We emphasize that this work is not intended to comment on the relationship between these components, or the details of how the transition from pop-III to population-II star formation occurs, but only to examine some possible scenarios for the r-EBL, of which we presently have only indirect evidence.

\subsection{Evolution of background photons}

In general, the EBL in place at a given redshift $z_0$ can be calculated by an integral over source terms at all redshifts $z>z_0$, with additional terms describing the effect of redshift on photon wavelength and flux redshift (e.g., \citealp{sgpd11}).
\begin{equation}
J_\nu(\nu_0,z_0)=\frac{1}{4\pi} \int^{\infty}_{z_0} \frac{dl}{dz} \frac{(1+z_0)^3}{(1+z)^3}\epsilon (\nu,z) \, dz,
\label{eq:eblint}
\end{equation}
\noindent  where $\epsilon(\nu,z)$ is the galaxy emissivity at redshift $z$ and frequency $\nu=\nu_0(1+z)/(1+z_0)$, and $dl/dz$ is the cosmological line element, defined as
\begin{equation}
\frac{dl}{dz}=\frac{c}{(1+z)H_0} \frac{1}{\sqrt{\Omega_m(1+z)^3+\Omega_\Lambda}}
\label{eq:cosline}
\end{equation}
for a flat $\Lambda$CDM universe \citep{peebles93}.
An EBL component originating entirely above a redshift $z_r$ will evolve only through passive redshifting at $z<z_r$, with an invariant comoving photon number density.  If this component is observed locally with a spectral energy distribution $J'_\nu(\nu_0)$, then the proper flux of these photons at redshift $z<z_r$ is
\begin{equation}
J'_{\nu}(\nu,z)=(z+1)^3 J'_\nu(\nu_0); \;  \nu_0=\nu/(z+1)
\end{equation}

If $J'_{\nu}$ refers in this case to the EBL produced by early low metallicity stars, and $z_r$ is the end of the era of preeminence of these sources, then it is easy to see that $J'_{\nu}(z)$ will become an increasingly large fraction of the total background as $z$ approaches $z_r$.  This is due to the fact that the majority of the background light emerging from resolved galaxies after reionization in recent models (e.g. \citealp{sgpd11,dominguez11,franceschini08}) comes from redshifts considerably lower than $z_r$.  In the fiducial model of \citet{sgpd11}, for instance, $\sim 75$ per cent of p-EBL photons between 0.1 and 10 microns are emitted at $0 < z <2$, and 96 per cent at $z<4$.

\subsection{Gamma-ray attenuation}

Gamma-gamma scattering into electron-positron pairs can occur when there is sufficient energy in the center-of-mass frame of the two-photon system.  Including the effect of interaction angle as measured in the cosmological frame, this condition is
\begin{equation}
\sqrt{2E_1 E_2 (1-\cos\theta)} \geq 2 m_e c^2,
\label{eq:paircre}
\end{equation}
where $E_1$ and $E_2$ are the photon energies and $\theta$ is the angle of incidence, which for our purposes is a random distribution on the unit sphere.    We can rewrite Equation \ref{eq:paircre} to define the minimum threshold energy $E_{th}$ for a background photon to interact with a gamma ray of energy $E_{\gamma}$,
\begin{equation}
E_{th}=\frac{2m_e^2c^4}{E_\gamma (1-\cos\theta)}.
\label{eq:threshold}
\end{equation}
\noindent The cross-section for this process is
\citep{breit&wheeler34,gould&schreder67,madau&phinney96}
\begin{align}
 \sigma(E_1,E_2,\theta) & =  \frac{3\sigma_T}{16}(1-\beta^2) \\ & \times  \left[ 2\beta(\beta^2-2)+ (3-\beta^4)\ln \left( \frac{1+\beta}{1-\beta}\right)\right], \nonumber
\label{eq:sigma}
\end{align}
where
\begin{equation}
\beta = \sqrt{1-\frac{2m_e^2c^4}{E_1 E_2 (1-\cos\theta)}},
\end{equation}
and $\sigma_T$ is the Thomson scattering cross section.  We can also revisit Eq. \ref{eq:threshold} to define the minimum energy or maximum wavelength for the background photons that can interact with a gamma ray of energy $E_\gamma$ as
\begin{equation}
E_{bg}=\frac{m_e^2 c^4}{E_\gamma} = 0.261\; \left(\frac{\mbox{TeV}}{E_\gamma}\right) \;\mbox{eV},
\end{equation}
\noindent or equivalently,
\begin{equation}
\lambda_{bg}=475  \: \left(\frac{E_\gamma}{10 \; \mbox{GeV}}\right) \; \mbox{\AA}.
\label{eq:charwave}
\end{equation}
These equations all refer to rest-frame energies.

To calculate the optical depth for a gamma ray observed at energy $E_\gamma$, we perform the integral along the line of sight to the target at redshift z \citep{stecker92},  
\begin{align}
\label{eq:opdep}
& \tau(E_\gamma,z_0) =  \frac{1}{2}\int^{z_0}_0 dz\;\frac{dl}{dz}\int^1_{-1}du \; (1-u)\\  & \times  \int^{\infty}_{E_{min}} dE_{bg}\; n(E_{bg},z)\;\sigma(E_\gamma (1+z),E_{bg},\theta). \nonumber
\label{eq:opdep}
\end{align}
\noindent Where we have \[E_{min}=E_{th}\:(1+z)^{-1}=\frac{2m_e^2c^4}{E_\gamma (1+z)(1-\cos\theta)}\] to account for the redshifting of the gamma-ray energy.  Here $n(E_{bg},z)$ is the proper density of target background photons as a function of energy $E_{bg}$ and redshift $z$, and $u$ is shorthand for $\cos\theta$.  $dl/dz$ is the cosmological line element, Equation \ref{eq:cosline}.  

The proper photon number density of a passively evolving component increases as $(z+1)^3$, and the gamma-ray attenuation per unit proper distance at a given energy therefore increases by the same ratio.  We can place the strongest constraints on reionization-era fields using the highest redshift gamma-ray sources available, provided that these gamma rays satisfy the minimal energy condition discussed above.

\subsection{Spectral model}
\label{sec:spectmod}
Several emission processes can contribute to the cosmological emissivity at the time of reionization.  The fact that we are interested only in the broadband features of the spectrum, with integration over a range in redshift, will allow us to make some simplifying assumptions.   A simple description of the stellar spectrum at non-ionizing wavelengths as a thermal blackbody is sufficient for massive pop-III stars that emit a large portion of their energy at ionizing wavelengths \citep{santos02}.  Ionizing radiation will undergo processing by neutral hydrogen within the host halo and eventually emerge at longer wavelengths, and the fraction of emission that leaves the host halo (denoted by the escape fraction $f_{esc}$) will be similarly affected by the neutral hydrogen residing the intergalactic medium (IGM).

In this work, we focus on the contribution of stars with masses from 5 to 500 \msun~to the cosmological background radiation. For this purpose, we create rest-frame SED templates that include the stellar contribution as a truncated blackbody, plus \lya, free-bound, and two-photon emission from reprocessed ionizing radiation.  The mean free path for ionizing radiation is assumed to be short at all wavelengths we consider, and our templates have no emission above the Lyman limit.  We ignore the contribution to the spectrum from free-free interactions.  While important at longer wavelengths, these only account for a small fraction of the emission at optical and UV wavelengths that are the main concern in this work.  Likewise we ignore the contribution of other recombination lines, which have only a relatively small effect that occurs at longer wavelengths.  While these templates are not intended to accurately represent fine spectral features, they do provide a reasonable estimate of the energy released into the cosmological background photon fields by early stars at rest-frame wavelengths from the Lyman limit to the optical, and can be raised or lowered to describe the output from a given global star-formation history.

The IMF of low- and zero- metallicity stars is highly uncertain.   Therefore, we consider two different possible IMFs in this work.  The first is a simple Salpeter-like power law function from 5 to 500 \msun, while the second is a much more top-heavy  Larson \citep{larson98} IMF, defined as 
\begin{equation}
\frac{dN}{dM} = M^{-1}(1+M/M_{c})^{-1.35},
\label{eq:larson}
\end{equation}
which has a Salpeter slope at high masses ($>M_c$) and a flat slope at lower masses, reducing the number of low-mass stars.  Following \citet{fernandez10}, and to allow a comparison with them in the next section, we set $M_{c}=250$ \msun.  Our investigation of two different possibilities is motivated by the suggestion from simulations that there is a critical metallicity below which the formation of low-mass stars is strongly inhibited, leading to an IMF that is considerably more top-heavy than later population II stars \citep{tumlinson06,norman08}.  Whether this leads exclusively to very high mass stars (e.g. $>100$ \msun), and/or to a significant departure from a Salpeter-like slope in the high mass IMF ($dN/dM \propto M^{-2.35}$, \citealp{salpeter55}) is disputed by various authors \citep{bromm02,bromm&loeb04,tan&mckee04,tumlinson04}.  

For the stellar component we apply a blackbody spectrum down to a wavelength 912 \AA.  Descriptions of the spectral temperatures, lifetimes, and number of ionizing photons released by pop-III stars are taken from \citet{schaerer02}.  As noted here and in \citet{fernandez&komatsu06}, the collective emission properties of high-mass stars {\em per stellar mass} are highly insensitive to mass above a certain threshold, $\gtrsim 50 - 100$ \msun.  Above this threshold, stars are well described by an effective temperature $\sim 10^5$ K, and release $\sim 10^{62}$ hydrogen-ionizing photons per solar mass over a lifetime.  In the regime between the 5 \msun~cutoff and this mass scale a dependence on the IMF still exists.   In the case of our Salpeter IMF, about 81 per cent of mass in a zero-age population exists in stars of less than 100 \msun, and about half in stars less than 20 \msun, necessitating consideration of this deviation from the high-mass properties.  In the case of the Larson IMF, $<100$ \msun~stars contain only about one third of all stellar mass.  Integrated properties with this IMF are therefore closer to what would be expected of an extreme IMF, populated only by $>100$ \msun~stars.

The Lyman-$\alpha$ line is the most significant spectral feature expected to appear in the reionization-era background.  While higher-order Lyman-series lines are suppressed by repeated scattering event in the IGM, the photons in the $\alpha$ line will eventually escape out of resonance due to redshifting and propagate freely through the universe \citep{loeb&rybicki99}.  This is true regardless of whether the conversion of ionizing radiation to \lya~occurs inside the host halo or outside in the IGM. The overall spectrum is therefore not significantly impacted by the ionizing escape fraction, which determines this ratio.   While varying the escape fraction can alter the shape of the angular perturbations in the near-IR EBL \citep{fernandez10},  the broadband properties of the UV emission are not affected by the details of this process \citep{fernandez&komatsu06}.  A simple assumption to determine the line strength is that some fraction of the energy emitted at ionizing wavelengths (which is excluded from the stellar spectrum) eventually emerges as \lya~emission \citep{santos02,fernandez&komatsu06}.  Following \citet{santos02}, Eq. (27), we assume 0.75 \lya~photons for each H-ionizing photon, plus an additional 0.53 for each He-II ionizing photon.  The latter occur at about half the rate of hydrogen ionizing photons, for a high-mass star.  

Free-bound and free-free continuum radiation account for a sub-dominant component of the emission in the UV.  We include a description of the free-bound contribution in our spectral model, as per the derivation in \citet{fernandez&komatsu06}.  As in that work, we assume an average gas temperature of $T=2 \times 10^4$ K.
\begin{align}
& \varepsilon_\nu (\nu,z) = 1.05 \times 10^{-34} \frac{dQ_{H}}{d(M_\odot)} \; \dot{\rho}_*(z) \frac{R}{kT} \times \\ &  \sum_{n=2}^{\infty} \frac{e^\frac{R}{kTn^2}}{n^3} \; g_{fb}(n) \; \phi_{2} (T) \; e^{-\frac{h\nu}{kT}} \nonumber \mbox{erg s$^{-1}$ Hz$^{-1}$ Mpc$^{-3}$}.
\end{align}
Here $ \dot{\rho}_*(z)$ is the star-formation rate density (SFRD) in units of \msolar~yr$^{-1}$ Mpc$^{-3}$, and $dQ_{H}/d(\mbox{M}_\odot)$ is the average number of ionizing photons released by the creation of 1 solar mass of stars for an assumed IMF.  $R$ = 13.6 eV, $g_{fb}$ is the free-bound Gaunt factor and $\phi_2(T)$ is the recombination coefficient to all states other than the ground state, as described in \citet{spitzer78}, Table 5.2.  Both of these are close to one for our purposes.  The summation over energy levels is truncated at n=3 in our calculation, as higher levels mainly contribute to longer wavelengths.

A certain number of recombination photons will descend to the ground state by two-photon emission rather than \lya.  As described in \citet{spitzer78}, two photon emission is required for de-excitation of an electron in the $2^2$S state, where single photon emission is prohibited.  The probability of this occurrence is a weak function of temperature, and we use the value of $1-z_{L\alpha}=0.36$ listed in Table 9.1 of \citet{spitzer78}.  The emitted spectrum, in terms of photon number, necessarily forms a symmetrical peak around 1/2 the \lya~energy (2432 \AA~or $\nu_{2\gamma} = \nu_{\mbox{\scriptsize{Ly}}\alpha}/2$); the majority of the two-photon flux emerges at UV wavelengths.  The fitting function for the spectrum of two-photon emission used here is taken from \citet{brown&mathews70}:
\begin{align}
& \varepsilon_\nu(\nu)\propto \frac{\nu}{nu_{2\gamma}}\lbrack1.307-2.627\left(\frac{\nu-\nu_{2\gamma}}{2\nu_{2\gamma}}\right)^2 \\ & +2.563\left(\frac{\nu-\nu_{2\gamma}}{2\nu_{2\gamma}}\right)^4-51.69\left(\frac{\nu-\nu_{2\gamma}}{2\nu_{2\gamma}}\right)^6\rbrack. \nonumber
\end{align}

The total emissivity, which is then the input $\epsilon (\nu,z) $ for Eq.~\ref{eq:eblint}, is the sum of the \lya~, stellar, free-bound scattering, and two-photon contributions.  We assume an instantaneous approximation for the emission of radiation, i.e., all light from a star is emitted the instant the star is created.  Since the lifetime of high mass stars is very short compared to cosmological timescales (6.2 $\times 10^7$ yr for a 5 \msolar~star, 1.04 $\times 10^7$ yr for a 15 \msolar~star; \citealp{schaerer02}), this is reasonable approximation that will have little impact on our results.

\section{Results}
\label{sec:results}

In \S \ref{sec:thermlims}, we begin by setting some general limits on the pop-III contribution to the present-day EBL.  This will be done by considering the highest energy photons detected from two distant gamma ray sources, and making the simple requirement that a passively evolving photon population not create an optical depth that is greater than 1.  In \S \ref{sec:highzsf}, we will address the how these limits on the EBL can be translated into bounds on pop-III star formation.

\subsection{Limits on the EBL flux from reionization}
\label{sec:thermlims}

\begin{figure}
\includegraphics[width=\columnwidth]{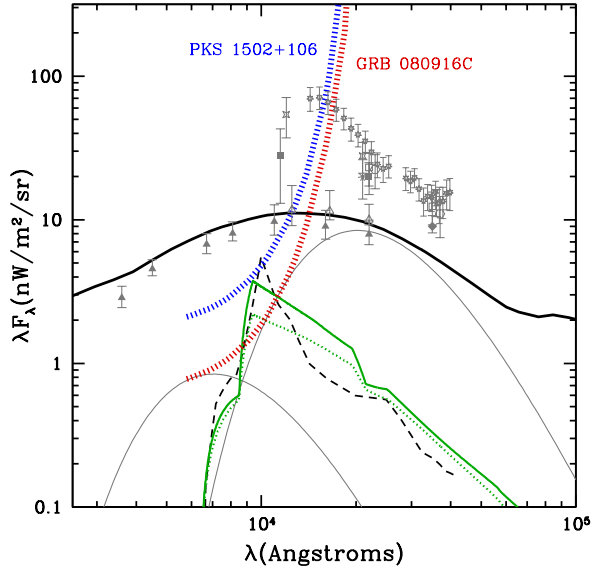}
\caption{ Upper limits placed on the contribution to the r-EBL by the assumption of the optical thinness of the universe to the highest energy photons seen by {\em Fermi} from GRB 080916C (red, \citealp{abdo09a}) and PKS 1502+106 (blue, \citealp{abdo10b}).  The solid and dotted green lines are the SEDs of the p-EBL produced by pop-III stars forming with a Larson and Salpeter IMF, respectively, and with SFRD of 0.2 \msolar~yr$^{-1}$ Mpc$^{-3}$ terminating at $z_r=6$.  For comparison, the Larson spectrum with a star-formation efficiency of $f_*=0.01$ from \citet{fernandez10} is shown as well (dashed black line).  The two thin grey curves are examples of the thermal spectra used to create the optical thinness r-EBL bound; here they are shown tangent to the limit for GRB 080916C.
The thick black line is the predicted p-EBL from the fiducial model of \citet{sgpd11}.  Other grey points are total EBL measurements, including the lower limits from number counts from \citet{madau00} and \citet{keenan10} (solid and open upward-pointing arrows), and sky photometry measurements of DIRBE with sky subtraction:  \citet[][solid squares]{wright01}, \citet[][crosses]{cambresy01}, \citet[][solid diamond]{levenson08}, \citet[][open hexagons]{gorjian00}, \citet[][open squares]{wright&reese00}, and \citet[][asterisks]{levenson07}. The small stars are direct measurements from the IRTS experiment \citep{matsumoto05}.  Some points are slightly offset to improve readability.
}
\label{fig:ebllims}
\end{figure}

We find that a simple gamma-ray optical thinness criterion can significantly limit the high-redshift contribution to the observed EBL.  In Fig.~\ref{fig:ebllims}, we show the limits placed on EBL contributions from high redshift, based on two different gamma-ray sources that have been recently observed with {\em Fermi} LAT.  The first, GRB 080916C \citep{abdo09a} was a bright gamma-ray burst with a measured redshift of z=4.35 \citep{greiner09}, which was observed by {\em Fermi} LAT and GBM on September 16, 2008.  The spectrum of the source at high energy was found to be continuous over nearly 6 orders of magnitude, and the highest energy event was a 13.6 GeV photon, corresponding to a rest-frame energy of 73 GeV.  The other source is flat-spectrum radio quasar PKS 1502+106, which was detected during a flaring event at z=1.839 \citep{abdo10b}.  The spectrum of this source was found to be most satisfactorily fit by a log-parabolic function peaking near 1 GeV.  \citet{fermiEBL} reported a highest energy photon of 48.9 GeV from this source, corresponding to a rest-frame energy of 139 GeV.  

The thick colored lines in the figure delineate the allowed present-day flux as a function of wavelength that does not produce an optical depth greater than 1 for the highest energy photons from each of these sources.  These contours were created by considering a large number of hypothetical r-EBL SEDs, and excluding those for which the optical depth exceeded unity.  These results are sensitive to the nature of the SED types considered - for this exercise we have used blackbody spectra at a variety of temperatures and normalizations.  Because the optical depth from pair-production interactions (Eq. \ref{eq:opdep}) depends on the integrated spectrum falling within the allowed energy range, the actual bound  is dependent on the assumed type of spectral feature; a more sharply-peaked SED would be less constraining.  Our use of simple blackbody spectra avoids the introduction of any parameters into the results at this point.
A couple of sample blackbody spectra are shown on this plot, each at a normalization that would produce optical depth unity for the highest energy photon detected from GRB 080916C.

This plot also shows the r-EBL SED models described in \S \ref{sec:spectmod} that we employ in the next section, assuming a star formation rate density (SFRD) of 0.2 \msolar~yr$^{-1}$ Mpc$^{-3}$ from $z=15$ to $z=6$.  We have also plotted the EBL prediction from \citet{fernandez10}, when using a star-formation efficiency of $f_* = 0.01$.  Differences in the shape of the prominent peak created by \lya~emission are due to the fact that star-formation is not constant in their model, and decreases with redshift.

\begin{figure}
\includegraphics[width=\columnwidth]{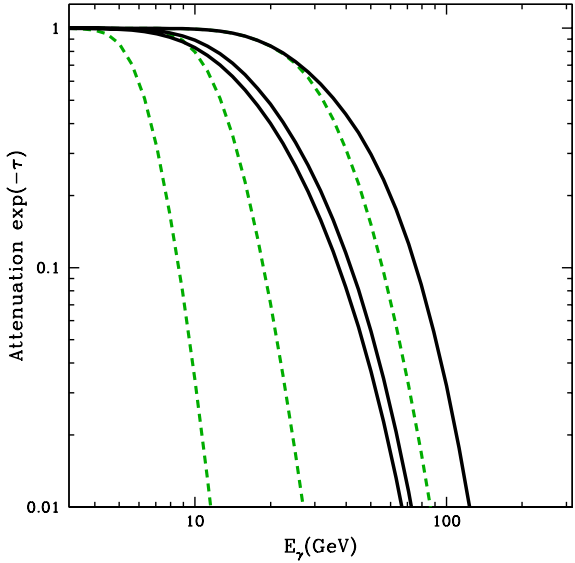}
\caption{Attenuation for gamma-rays from high redshift sources as a function of observed energy.  Solid black lines show the attenuation ($e^{-\tau}$) to gamma rays from sources at $z=$ 2, 4, and 6 (upper-right to lower left) predicted by the fiducial model of GSPD11 (p-EBL; solid black line in Fig. \ref{fig:ebllims}).  Dashed green curves are the attenuation for sources at these same redshifts including the r-EBL component from Fig. \ref{fig:ebllims}, originating from high redshift pop-III stars with a Larson IMF and SFRD 0.2 \msolar~yr$^{-1}$ Mpc$^{-3}$, ending at $z_r=6$. }
\label{fig:opdep}
\end{figure}

Our results here demonstrate the power of high redshift gamma-ray source observations in limiting flux contributions from high redshift.  Both results constrain the flux at $\lambda \lesssim 1.4$ \mum~from reionization to be a fraction of that produced below the redshifts of the two sources.  Fig.~\ref{fig:opdep} demonstrates the large increase in optical depth that the r-EBL components can create at high redshift compared to the p-EBL predictions in a typical model.  By redshift 4, the Larson pop-III contribution shown in Fig. \ref{fig:ebllims} produces a spectral cutoff at substantially lower energy than the GSPD11 p-EBL alone, despite the fact that the latter dominates the former at all wavelengths in the $z=0$ EBL.

\subsection{High-redshift star formation}
\label{sec:highzsf}
By linking gamma-ray opacity to a particular high-redshift emission scenario, as described in \S \ref{sec:spectmod}, we can constrain the amount of star formation consistent with a given set of gamma-ray observations.  The limits derived will necessarily be dependent on the IMF applied, and the redshift at which zero-metallicity star-formation is assumed to terminate.

\subsubsection{The highest-energy photon method}

In this section, we describe how spectra from high-redshift gamma-ray sources observed by the {\em Fermi} LAT can be used to set limits on the background placed by pop-III stars.  Because the most constraining sources will be those at the highest redshifts, we are interested in the small number of AGN and GRBs that have been detected at $z \gtrsim 2$ by {\em Fermi}.  A list of `most constraining' sources was provided in \citet{fermiEBL}.  To determine if a particular emission scenario is allowed or prohibited, we apply the highest energy photon method described in that paper.  In this method, a power-law $dN/dE=A\, E^{-\Gamma}$ is fit to the lower-energy signal from the source, which is assumed to be unaffected by photon-photon interactions (for PKS 1502+106, a log-normal parametrization was used).  Once the parameters $A$ and $\Gamma$ are determined, the opacity induced by the background photon field, together with instrument parameters, are used to determine the probability distribution for the highest energy photon detected.  If the likelihood of detecting this photon at or above the energy of the observed highest energy photon ($E_{high}$) is sufficiently small, then the opacity model can be excluded with a defined confidence level.  This method provides a more robust exclusion of background models than the alternative of simply using a predefined value in gamma-ray opacity $\tau$ at $E_{high}$, as we have done in Fig.~\ref{fig:ebllims}, because it takes into consideration the normalization of the spectrum and the predicted number of observable gamma rays at and above $E_{high}$ in the absence of EBL opacity, which may be much smaller or larger than unity.

The expected number of photon counts above energy $E_{high}$ is
\begin{equation}
N_x(>E_{high})=A \; \int^{\infty}_{E_{high}} C(E)\, E^{-\Gamma}e^{-\tau(E)} dE
\label{eq:nex}
\end{equation}
Here $A$ and $\Gamma$ are the fit constants as previously described, while $\tau(E)$ is the gamma-ray optical depth for the source at observed energy $E$, and $C(E)$ is an energy-dependent factor describing the exposure of the source with {\em Fermi} LAT, i.e., instrument effective area integrated over time.  The probability of observing zero counts is then described from Poisson statistics as 
\begin{equation}
P(0)=e^{-N_x(E_{high})}.
\end{equation}

We have analyzed 5 high-redshift {\em Fermi} LAT sources selected from the `most constraining' list of \citet{fermiEBL}, using publicly available data and software tools from the {\em Fermi} Science Support Center.\footnote{http://fermi.gsfc.nasa.gov/ssc/}  The properties of the sources are summarized in Table \ref{tab:latsources}.  Our analysis uses the `gtltcube' routine to determine source exposure, and the `gtpsf' function to calculate the energy-dependent point-spread function (PSF).  We use the P6\_V3 instrument response function, which is the default at the time of writing.  Integrated data from the full instrument lifetime is used for the AGN sources, and from $T_0 -5$ to $T_0+135$ s for GRB 090816C; $T_0$ being the trigger time of the burst in the GBM instrument.  Note that 135 s is nearly 3 times the interval of GeV emission from the GRB presented in \citet{abdo09a}.

\begin{table*}
\centering
\begin{tabular}{@{}lccccccc}
  \hline
Source ID &  Redshift & Spectral Index & $N_{norm}$ & $E_{high}$ (GeV)  & $\tau_{fid}$ & $N_{x,fid}(>E_{high})$ & $P_{pEBL}(\geq 1)$  \\
\hline
GRB 080916C & 4.35 & -2.15 & 36 & 13.2 & 0.34 & 0.35 & 0.295 \\
\hline
PKS0227-369 & 2.11 & -2.56 & 146 & 31.9 & 0.66 & 0.20 & 0.18 \\
\hline
PKS1502+106 & 1.84 & -2.36 & 2223 & 49.2 & 1.07 & 1.06 & 0.65 \\
\hline
PKS0805-07 & 1.84 & -2.09 & 973 & 46.7 & 0.96 & 2.4 & 0.91 \\
\hline
J1016+0513 & 1.71 & -2.27 & 322 & 46.7 & 0.83 & 0.52 & 0.41 \\
 \hline
 \end{tabular}
 \caption{Statistics for the five gamma-ray sources we consider in constraining pop-III star formation.  Columns show the source name, redshift, and spectral index.  $N_{norm}$ refers to the number of photons falling between 1 and 10 GeV (0.5 and 5 for GRB 0809156C) that were used to determine the normalization of the source spectrum.  $E_{high}$ is the highest energy photon received from the source, and $\tau_{fid}$ is the optical depth of that photon produced by the p-EBL predicted in GSPD11. $N_{x,fid}(>E_{high})$ is the expected number of photon counts at and above $E_{high}$ after applying the optical depths from the GSPD11 post-reionization EBL, but before considering any contribution from pop-III stars, and $P_{pEBL}(\geq 1)$ is the probability of detecting at least one photon above $E_{high}$.}
 \label{tab:latsources}

\end{table*}

The photons within the 95 per cent containment region of the PSF around the AGN source in each energy bin are then assumed to be associated with the source.  For GRB 080916C, the 99 per cent containment region is used.  To determine the normalization parameter ($A$) in Equation \ref{eq:nex} above, we use counts between energies of 500 MeV and 5 GeV for GRB 080916C, and 1 and 10 GeV for the four AGN.  

\subsubsection{Combined probabilities method}
\label{sec:comb}

An alternative method is to consider the combined probabilities of having $>1$ photons at or above the highest observed photon energy for each source, to put a stronger constraint on background fields.  That is, we consider the combined probability 
\begin{equation}
P_{tot}(\geq 1)=\prod_i P_i(\geq 1) 
\label{eq:ptotunnorm}
\end{equation}
of detecting photons from all sources considered, and exclude background scenarios that lead to a $P_{tot}$ less than a given value.  
However, a potential pitfall of this method is that $P_{tot}$ can be considerably less than 1 even in the absence of a pop-III contribution; due to the impact of the p-EBL, which is quite uncertain at high redshifts.  As discussed in G09 and GSPD11, this uncertainty in the UV background is a factor of several at $z \gtrsim 2$.  For high determinations of the UV background, like the fiducial model in GSPD11, the optical depth for the highest energy photons can be as high as 1, as shown in Table \ref{tab:latsources}.  The p-EBL effect is therefore not something that can be ignored here.  Result derived using Eq. \ref{eq:ptotunnorm} above will be necessarily be a strong function of the assumed star-formation model.  
In Table \ref{tab:latsources}, we also show the expected number of photons expected above $E_{high}$ for each source, and the associated probability of detecting at least one photon at or above the highest photon energy.  

As we are interested in isolating the impact of the r-EBL, we first compensate for the impact of the p-EBL on our results by renormalizing the combined probability $P_{tot}$ to 1, after convolving the observed spectrum with a given p-EBL opacity. Thus we define
\begin{equation}
P_{tot}(\geq 1)=\frac{\prod_i P_i(\geq 1)}{\prod_i P_{i,pEBL}(\geq 1)},
\label{eq:combprop}
\end{equation}
where $P_{i,pEBL} (\geq 1$) is the probability of 1 or more photons from a source, after considering only an assumed p-EBL model, while $P_i(\geq 1)$ considers both background components.  This change effectively isolates the impact of the r-EBL on photon detection probability.  The resulting limits on the reionization-era background are then only a weak function of the p-EBL;  we find a variance of only about 10 per cent in our results for the pop-III star formation rate limits if the p-EBL flux is increased or decreased by a factor of 2.  The following results assume the p-EBL predicted by the fiducial model of GSPD11.

\begin{figure}
\includegraphics[width=\columnwidth]{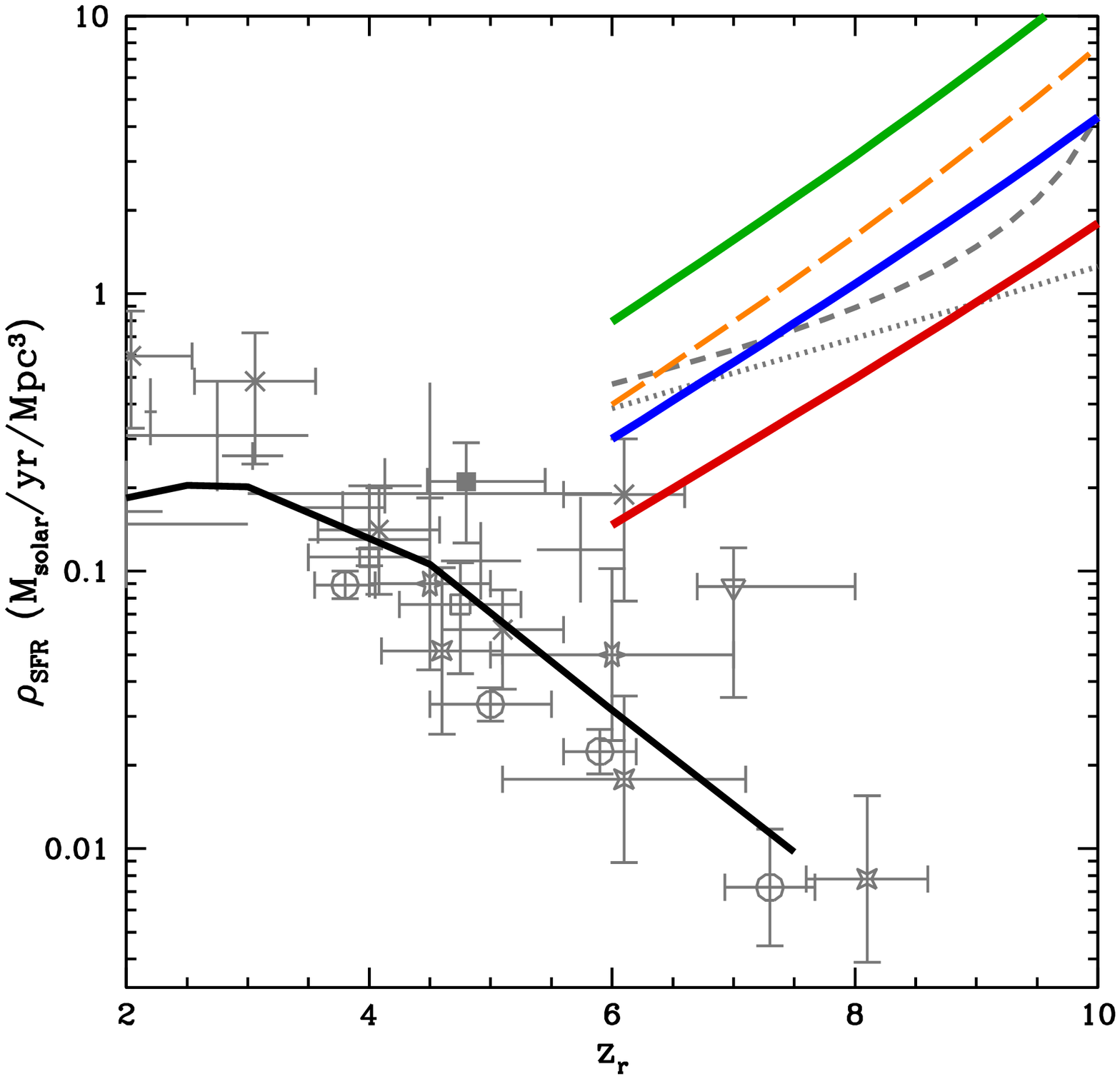}
\caption{Upper limits placed on pop-III star formation rate density by our analysis of the spectra of high-redshift {\em Fermi} LAT detections.  These results are for spectral output from stars assuming a Larson IMF.  Solid red, blue, and green lines (bottom to top on right-hand side of plot) are limits set on SFRD for significance values of 2$\sigma$, 3$\sigma$, and 5$\sigma$ for the combined analysis (Eq.~\ref{eq:combprop}) of all 5 sources listed in Table \ref{tab:latsources}.  These correspond to probabilities of 0.046, 0.0027, and $6.0 \times 10^{-7}$ of finding the observed pattern of highest energy photons across all sources.  The long-dashed orange line shows the 2$\sigma$ bound for the analysis of GRB 080916C, when considered alone.  The dotted grey line at $z \geq 6$ shows the star formation rate level that would enrich the IGM to 0.1 $Z_\odot$ by $z_r$, while the dashed grey line corresponds to a star formation efficiency of $f_* =0.1$ in collapsed structures; see text for more details on these criteria.  We have also shown SFRD measurements at a variety of redshift.  Because these measurements presumably include cooler pop-II stars at masses less than our cutoff of 5 \msolar, they are not directly subject to our proposed limits on pop-III stars, but are included here for comparison.  The solid black line is the star-formation history predicted in the semi-analytic model of \citet{sgpd11}; this has been converted from a Chabrier IMF \citep{chabrier03} to a Salpeter with a multiplicative factor of 1.6, which is typical of the conversions described in \citet{wilkins08}.  Grey error bars without symbols are from the compilation of \citet{hopkins&beacom06}.  Others measurements based on redshifted optical/UV light include \citet{bouwens07} (open circles), \citet{verma07} (solid squares), \citet{ouchi04} (open squares), the upper limit from \citet{mannucci07} (downward-pointing triangle), and \citet{thompson06} (crosses).  The latter of these has been converted from a Chabrier IMF  to Salpeter in the same manner as the semi-analytic model.  Star-like symbols are results from GRB rate analyses: \citet{wang&dai09} (4-stars) and \citet{yuksel08} (6-stars).  Some points have been shifted slightly for readability.}
\label{fig:sfrd_lar}
\end{figure}

\begin{figure}
\includegraphics[width=\columnwidth]{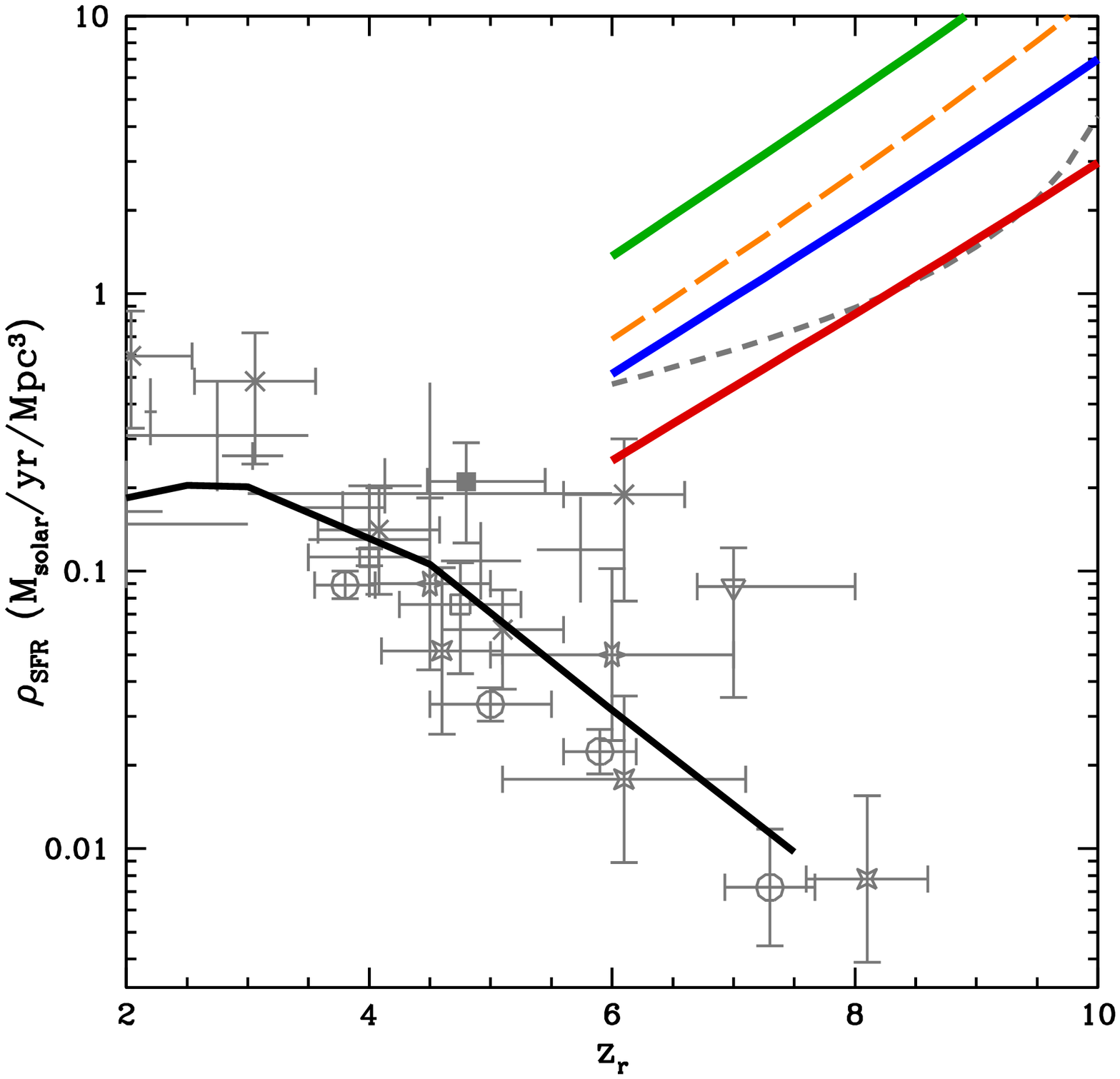}
\caption{As in the previous plot, but showing results for a Salpeter IMF ($dN/dM \propto M^{-2.35}$).  Limits on star formation from IGM enrichment due to pair-instability supernovae are more than an order of magnitude higher in this case, and are not shown here.}
\label{fig:sfrd_sal}
\end{figure}

In Figs.~\ref{fig:sfrd_lar} and \ref{fig:sfrd_sal}, we show the overall limits on SFRD obtained using both the combined model, and a singular analysis using only GRB 080916C.  Singular analyses with the other sources always lead to comparatively weak upper bounds, $\gtrsim 1$ \msolar~yr$^{-1}$ Mpc$^{-3}$ at the $2\sigma$ level.  A collection of observational data, as well as the model results for SGPD11 are shown as well; however most of these measurements reflect the luminosity of the brightest sources seen at these epochs, and are generally assumed to be due to population I/II stars in environments too metal-rich to support pop-III star formation.  The measurement based on observed GRB rates are free from this bias, however they suffer from their own systematic uncertainties \citep{kistler09,beckman10}.   The pop-III SFRD in our model is assumed to be constant in time from redshift 15 to the indicated cutoff redshift $z_r$.  The choice of a functional form for the SFRD history is necessary and arbitrary here, and we have chosen a constant history for simplicity.  In practice, the strongest effect on gamma-ray opacity will come from photons emitted near $z_r$, as evidenced by the rapid increase in SFRD upper limits with increasing $z_r$ in these figures.   Results in these figures can therefore be considered as approximate SFRD limits at $z_r$.  These limits are necessarily conservative, due to the renormalization that we have made in Eq. \ref{eq:combprop} that limits the influence of the poorly-constrained p-EBL on our results.  If the p-EBL contribution to the UV is high, then the actual limits on pop-III stars after considering the total background photon population would be stronger than what is depicted in the figures.

We have also shown in Fig.~\ref{fig:sfrd_lar} the limits on pop-III growth that exist due to constraints on global metallicity and baryons available in collapsed structures.  Pair-instability supernovae (\snpi) are responsible for a large release of metals into the IGM and ISM, which can potentially lead to conflict with the measured enrichment in high-redshift \lya-forest observations (e.g. \citealp{schaye03}).  As a very simple enrichment model, we assume that all massive stars with 140 \msolar~$<$ M $<$ 260 \msolar~undergo \snpi~upon death, and release half their mass into the IGM as metals \citep{portinari98}.  Enrichment from lower mass stars is ignored.  For our Larson IMF, about 9 per cent of all stellar mass falls in the \snpi~range, leading to a conversion of 4.5 per cent of stellar mass into IGM metals with a single generation of stars.  In Fig.~\ref{fig:sfrd_lar}, we show the level of global star-formation that between $z=15$ and $z=z_r$ would enrich the IGM to one-tenth solar metallicity.  This is already considerably higher than measurements such as those of \citet{aguirre02} and \citet{bouche07}, which suggest a median enrichment of $\lesssim 10^{-2}\: Z_\odot$ at redshift 2, so our bound is quite conservative.  There are several ways in which this metal production could be suppressed.  If the IMF were modified from the Larson function assumed here, so as to reduce the number of stars forming in the \snpi~mass range, metal production would be greatly diminished.  Stars at masses higher than $\sim 260$\msolar~are generally believed to follow a different evolutionary path than lighter stars and end with most of their metals inside remnant black holes \citep{heger&woosley02}.  Shifting high-mass star production to either above or immediately below the \snpi~mass range (i.e. 50 to 140 \msolar, or $\gtrsim$260 \msolar) would reduce IGM enrichment, while having little impact on the UV spectra produced per stellar mass.  It is also possible that our assumption of all metals escaping to the IGM after an \snpi~event is incorrect.  With a Salpeter IMF, the mass fraction of baryons allotted to the \snpi~mass range is much smaller, $\sim$0.6 per cent, and the bound would thus be a factor of 15 higher.

The number of baryons available at a given redshift to form stars puts another constraint on any early star-formation model.  The utilization of gas in collapsed structures can be quantified by the star-formation efficiency parameter $f_*$.  We take the total mass in all structures above the molecular cooling mass scale as a function of redshift from fig. 1 of MS05, and use eq. (6) of that paper to compute the total mass conversion into pop-III stars for a given value $f_*$; 0.1 is used here.  The correction factor $g$ in this formula is set to 0.65.  While slightly higher masses of $f_*$ might be allowable, MS05 argue that $f_* \gtrsim 0.3$ is probably implausibly high.

\section{Discussion and Conclusions}
\label{sec:disc}

Our results indicate that the GeV sources seen by {\em Fermi} at $z > 1.5$ disfavor a scenario in which pop-III stars with a strongly top-heavy IMF are formed in copious numbers (i.e. SFRD $\gtrsim 0.2$ to 0.4 \msolar~yr$^{-1}$ Mpc$^{-3}$) in the late stages of reionization, $6 < z < 8$.  At higher redshift, very high star-formation rate densities ($\gtrsim 1$) are disfavored by our result, and global limits on star-formation efficiency in proto-galaxies and limits on IGM metallicity impose a constraint at approximately the same level.  Switching to a more moderate IMF, i.e. a Salpeter IMF with a cutoff at 5 \msolar, raises our limits by about a factor of 1.7.

The results in Figs.~\ref{fig:sfrd_lar} and \ref{fig:sfrd_sal} can be compared to those calculated in RKM09 for zero-metallicity stars that are based on low-redshift blazar limits on the local background (their fig.~9; the model with $\alpha =10$ and $\beta = 0$ most closely resembles our simple step function applied for our pop-III SFRD function).  We place slightly stronger constraints on the SFRD at redshifts 7 and 8 than this work.  However, the significance of our strongest claims are only marginal, due to the very limited number of high-redshift photons that are relevant to our calculation.  The RKM09 result is based on an upper limit to the local background light of 5 nW m$^{-2}$ sr$^{-1}$ at 2 \mum.  As shown in Fig.~\ref{fig:ebllims}, our results most strongly constrain the EBL below $\sim 1$ \mum, and do not bound the EBL at co-moving wavelengths longer than 2 \mum.  For similar reasons, high redshift gamma-ray observations have little hope of providing meaningful constraints on a hypothetical contribution to the IR background from dark-matter burning stars \citep{freese08}.  Due to the relatively cool temperature of dark stars and the high redshifts at which they are theorized to exist, even an enormous contribution such as that in the maximal model of \citet{maurer10} would be at wavelengths too long for constraints to be derived, though this scenario would likely be difficult to reconcile with TeV-scale observations of lower-redshift blazars.  

We also strongly disfavor a scenario in which a near-IR excess proposed by \citet{matsumoto05} and a high EBL interpretation of the DIRBE 1.25 \mum~measurement (e.g. \citealp{cambresy01}) are produced by early stars.  Our results in Fig.~\ref{fig:ebllims} show that even a subdominant contribution to the EBL from high redshift at observed-frame wavelengths $\lesssim 1$ \mum~can be in disagreement with high energy data.  We find good agreement with the result of \citet{thompson07}, which disputes the detection of a high near-IR background, and attributes a possible contribution from high-redshift objects of 1-2  nW m$^{-2}$ sr$^{-1}$ at 1.4 to 1.8 \mum~using fluctuation analysis methods.

In an analysis of TeV blazars, \citet{orr11} recently found that the ratio of near- to mid-IR EBL flux needed to be significantly larger than proposed in most recent models, including those described in \citet{gspd11} and \citet{dominguez11}.  This work proposed a flux at 1.6 \mum~of $17 \pm 3$  nW m$^{-2}$ sr$^{-1}$, which is above the level produced by resolved galaxies is \citet{madau00} and at the upper 1$\sigma$ bound of \citet{keenan10}.  Our results do not strongly limit the possibility that an excess of a few nW m$^{-2}$ sr$^{-1}$ from high redshift could exist at or above this wavelength, but do disfavor a scenario in which such an excess continues below a wavelength of $\sim 1.2$ \mum.

\begin{figure}
\includegraphics[width=\columnwidth]{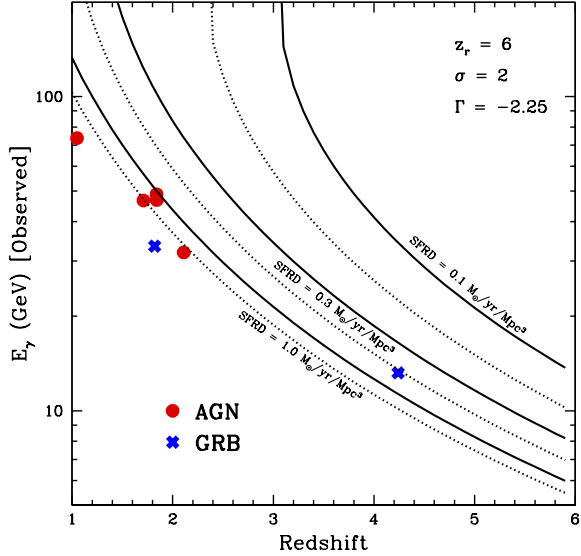}
\caption{Limits on high-redshift SFRD that could be possible with future gamma-ray observations.  Contour lines on the plot show the redshift and highest photon energy of a gamma-ray source that would exclude a SFRD above the level indicated at redshift 6 with 2$\sigma$ confidence.  We make the assumption here  that the gamma-ray source would have a spectral index $\Gamma=-2.25$ with no intrinsic curvature, and would have a normalization at lower energies such that one photon is expected at or above the highest observed energy in the absence of background fields.  Solid lines are results on SFRD for a Salpeter IMF, dotted lines are for a Larson IMF, and lines of each type are for limits of 1.0, 0.3, and 0.1  \msolar~yr$^{-1}$ Mpc$^{-3}$, from bottom left to upper right.  Symbols on  the plot indicate several of the most constraining sources observed thus far with the {\em Fermi} satellite.  In addition to the 5 sources in Table \ref{tab:latsources}, PKS 1144-379 and GRB 090902B are also shown.
}
\label{fig:sfrdcont_zr6}
\end{figure}

\begin{figure}
\includegraphics[width=\columnwidth]{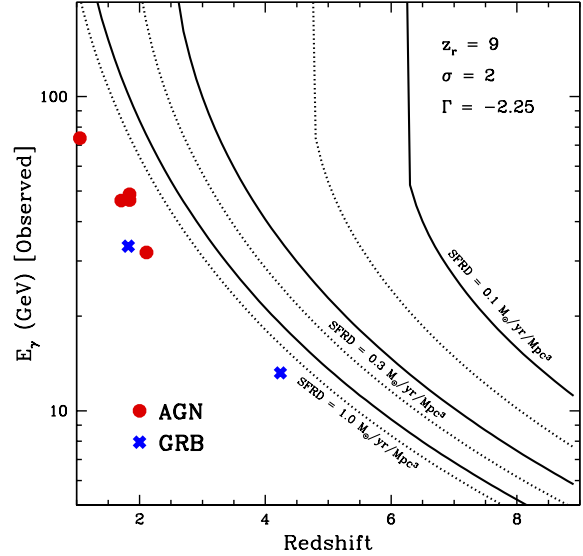}
\caption{As in the previous figure, but for a cutoff redshift $z_r = 9$.
}
\label{fig:sfrdcont_zr9}
\end{figure}

\begin{figure}
\includegraphics[width=\columnwidth]{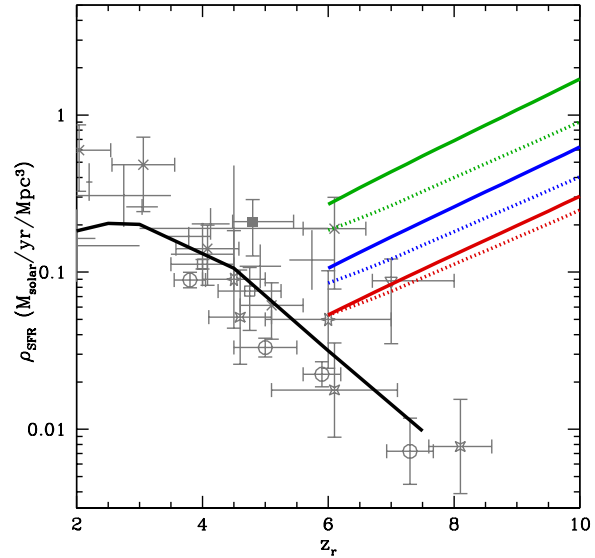}
\caption{Plot of the upper bounds on SFRD in two possible scenarios with future {\em Fermi} gamma-ray bursts, in the Larson IMF case.  The solid lines show the limits from a GRB with the same redshift and spectral characteristics of GRB 080916C, but with a highest energy observed photon of 30 GeV (160 GeV as emitted) instead of 13.2 GeV, in combination with the sources previously discussed.  The dotted lines show a case with a GRB at $z=7$, and a highest energy observed photon at 15 GeV (120 GeV emitted).  Line colors are as in Fig.~\ref{fig:sfrd_lar}.
}
\label{fig:sfrd_lar_higher}
\end{figure}

Our intent with this work is not just to compute the current limits on high-redshift star-formation available from gamma-ray observations, but also to demonstrate the usefulness of this method for future observations.  Figs.~\ref{fig:sfrdcont_zr6} and \ref{fig:sfrdcont_zr9} illustrate the SFRD limits that could be derived from future detections of high redshift sources with {\em Fermi} LAT or future telescopes.  In these plots, the axes refer to the redshift and highest observed photon energy $E_\gamma$ of a hypothetical gamma-ray source.  The source is then assumed to have a normalization at lower energy such that the expected number of photon counts at and above $E_\gamma$ is 1 [$N_x(>E_{high})=1$] in the absence of any background field.  The spectrum of the source is set here to -2.25, near the mean of the sources in Table \ref{tab:latsources}, and the p-EBL is ignored.  Given these parameters, the contours on the plots show the source redshift and $E_{\gamma}$ that would be required to place a given SFRD limit on pop-III star-formation at redshifts $z_r =6$ and 9, with 2$\sigma$ significance.  These contours are for limits derived based on a single source; combined limits for multiple sources like those in section \ref{sec:comb}, if available, would be somewhat stronger.  In Fig.~\ref{fig:sfrd_lar_higher}, limits based on two hypothetical high-redshift gamma-ray bursts are combined with the other sources of Table \ref{tab:latsources}.  This plot shows that new GeV sources, either at higher redshift than GRB 080916C, or at a similar redshift with higher energy emission, could strongly limit a pop-III contribution to star-formation in the late reionization period.

If the {\em Fermi} satellite remains in operation for its stated lifetime goal of ten years from its launch date, then its mission is currently less than one-third complete, and we can reasonably hope to see new GRB events or high-energy AGN photons that will strengthen our results.   The Cherenkov Telescope Array (CTA; \citealp{ctaconcept10}) is another possible source of constraining events.  CTA will have a lower threshold energy than current-generation ground-based instruments, and may be able to detect sources at much higher redshift than currently achieved from the ground.  Detections with either of these instruments could potentially shed new light on star-formation in the reionization era.

\section*{Acknowledgments}
RCG was supported during this work by a SISSA postdoctoral fellowship, and thanks W.~B.~Atwood, J.~Primack and A.~Bouvier for helpful discussions related to this project, and J.~Colucci  and the anonymous referee for reading the manuscript and providing useful comments.   Some calculations in the paper were performed on the SISSA High-Performance Computing Cluster.


\label{lastpage}
\end{document}